# Probing Mach's principle

Arto Annila[1,2,3,*]

[1]*Department of Physics,* [2]*Institute of Biotechnology and* [3]*Department of Biosciences, FI-00014 University of Helsinki, Finland*

**ABSTRACT**

The principle of least action in its original form á la Maupertuis is used to explain geodetic and frame-dragging precessions which are customarily accounted for a curved spacetime by general relativity. The obtained least-time equations of motion agree with observations and are also in concert with general relativity. Yet according to the least-time principle, gravitation does not relate to the mathematical metric of spacetime, but to a tangible energy density embodied by photons. The density of free space is in balance with the total mass of the Universe in accord with Planck's law. Likewise, a local photon density and its phase distribution are in balance with mass and charge distribution of a local body. Here gravitational force is understood as an energy density difference that will diminish when the oppositely polarized pairs of photons co-propagate from the energy-dense system of bodies to the energy-sparse system of the surrounding free space. Thus when the body changes its state of motion, the surrounding energy density must accommodate the change. The concurrent resistance in restructuring of the surroundings, ultimately involving the entire Universe, is known as inertia. The all-around propagating energy density couples everything with everything else in accord with Mach's principle.

## 1 INTRODUCTION

*The Science of Mechanics* written by Ernst Mach (Mach 1893) inspired Albert Einstein to question the notion of absolute space and time. Einstein reasoned that when everything depends on everything else, the overall distribution of matter in the Universe defines a metric tensor known as spacetime (Weinberg 1972; Berry 1974; Wheeler 1990). This fabric of cosmos is pictured to govern motions of bodies and trajectories of light. Notably the curved spacetime manifests itself also in rotational motion that does not appertain to the system itself. For example, the anomalous part of the perihelion precession of a planet does not ascribe to the solar system. Likewise, the axis of a gyroscope on board of a satellite does not stay aligned to a distant star but will gradually precess in the orbital plane as well as in the equatorial plane of a revolving central body. These geodetic and frame dragging drift rates were recently measured by Gravity Probe B (GP-B) orbiting the Earth (Everitt et al. 2011).

The GP-B data is in agreement with rotating-frame solutions of the Einstein field equations. Yet, the spacetime as a mathematical model is reticent in revealing the physical cause that imposes torque on the gyroscope's axis. Also in general, one remains puzzled what exactly is the mechanism how distant stars exert effects on local motions. To illuminate imprints of inertia (Heaviside 1893; Sciama 1953), we will examine perihelion, geodetic and frame-dragging precession by the principle of least action in its original form (Maupertuis 1744). When the universal law of nature is formulated as an equation of motion, its solutions are geodesics along which flows of energy propagate in the least time. This physical portrayal of space and time embodied by quanta on bound and free trajectories provides also a perspective to solutions of the Einstein field equations.

## 2 THE LEAST-ACTION PRINCIPLE

The variational principle in its original form states the conservation of total action so that a change in kinetic energy $d_t 2K$ balances changes in the scalar $\partial_t U$ and vector $\partial_t Q$ potentials (Maupertuis 1744; Sharma & Annila 2007; Annila 2009). For example, when considering celestial mechanics the balance

$$d_t 2K = -\partial_t U + \partial_t Q$$
$$\Rightarrow \partial_t mv^2 = -\partial_t \frac{GmM_\oplus}{r} + \partial_t mc^2 \qquad (1)$$

says that along the geodesic a change in the kinetic energy of a probe with mass $m$ in motion with velocity $v$ balances changes ($\partial_t = \mathbf{v} \cdot \nabla$) in the local gravitational potential due to a central mass $M_\oplus$ as well as in the universal gravitational potential. The squared speed of light $c^2 = GM/2R$ characterizes the potential due to the total mass $M = \int \rho 4\pi r^2 dr$ of the Universe, where $\rho = 1/2\pi Gt^2$ is the average density of matter and $G$ is the constant of gravitation (Unsöld & Baschek 2002; Koskela & Annila 2011). According to Eq. 1 combustion of quanta that are bound in matter to freely propagating photons at the lowest group of symmetry U(1) (Griffiths 1999; Annila 2010) has been powering the spontaneous breaking of densities at higher symmetry groups for the past $t = T \approx 13.7$ billion years. This on-going dilution of densities manifests itself as the expanding Universe of radius $R = cT$.

When Eq. 1 is re-expressed using the mass-energy equivalence $mc^2 = hf$, the energy density about the local body $M_\oplus$ can be placed relative to the universal surroundings in terms of the refractive index $n = c/v$

$$n^2 = \left(1 - \frac{GM_\oplus}{c^2 r}\right)^{-1} = (1-\phi)^{-1} \approx 1 + \frac{GM_\oplus}{c^2 r}. \quad (2)$$

The irrotational scalar potential $\phi$ due to the local density contributes to the density of free space defined as $n = 1$. Likewise, the surrounding density would adjust to any other local source, either still or in motion. For example, the energy density of space would revolve next to a rotationally flattened spheroid and adjust its phase density next to a charged body. Also a density that is confined by bodies would depart from the free, universal density.

## 3 MANIFESTATIONS OF LOCAL DENSITY

The local density at a radius $r$ from the body of mass $M_\oplus$ can be probed, e.g., by an orbiter with mass $m$. The local excess in density relative to the free space will be displayed in the orbiter's motions relative to distant stars. For example, the orbiter's path of radius $r$ will be longer by an arc $r\varphi$ in the excess density than in the free space. The added angle $\varphi$ per orbital period $\omega^{-1}$ can be determined from the least-time, i.e., stationary condition

$$d_t n^2 = 0 \Rightarrow d_t \left(\frac{r^2 \omega^2}{c^2} + \frac{GM_\oplus}{c^2 r}\right) = 0 \Rightarrow \omega^2 = \frac{GM_\oplus}{r^3}. \quad (3)$$

The resulting Kepler's third law can be used in the kinematic equation to express the excess $\Delta\theta = \theta_f(t) - \theta_i(0)$ between the initial $\theta_i(0)$ and final $\theta_f(t)$ angular position

$$\Delta\theta = \frac{1}{2}\alpha t^2 = \frac{1}{2}\frac{GM_\oplus}{r^3}t^2 = \frac{1}{2}\frac{GM_\oplus}{c^2 r}\frac{mc^2}{mv^2} = \frac{1}{2}\frac{GM_\oplus}{c^2 r}n^2 \quad (4)$$
$$= \frac{(2\pi)^2}{2}\frac{GM_\oplus}{c^2 r}\omega^{-2}\frac{mc^2}{mr^2} = (2\pi)^2 \frac{M_\oplus}{M}\frac{R}{r}\frac{mc^2}{I\omega^2} = \varphi \frac{mc^2}{I\omega^2}$$

where the angular acceleration $\alpha = GM_\oplus/r^3$ is used to give us the advancing angle $\varphi$ of the orbital axis as a fraction of the ratio of the universal potential $mc^2$ to the kinetic energy in the orbital motion $I\omega^2$ (Koskela & Annila 2011). The approximation of Eq. 2 implies that the advancing arc $r\varphi$ of the orbital axis will render itself measurable after the probe has completed numerous orbits. For example, peritellus precession of GP-B can be calculated to be 13.8 arc seconds per sidereal year (NASA Solar System Bodies, 2011). However, this would be difficult to determine precisely since the GP-B orbit is almost circular. The corresponding value of $\varphi$ for Mercury means that it will take a century to advance 43.1 arc seconds (Koskela & Annila 2011) in agreement with measurements (Clemence 1947). Despite the numerical consent the functional form of geodesic obtained by the principle of least action is distinct from that derived from general relativity.

According to Eq. 4 the satellite will sense, due to its orbital motion, an excess of density $GM_\oplus/2c^2r$ relative to the free space. The total excess of density relative to the universal potential $mc^2$ in terms of the refractive index

$$n^2 = \left(1 - \frac{GM_\oplus}{c^2 r} - \frac{GM_\oplus}{2c^2 r}\right)^{-1} \approx 1 + \frac{3}{2}\frac{GM_\oplus}{c^2 r} = 1 + \frac{mr^2\alpha}{mc^2} \quad (5)$$

yields the torque $\tau = I\alpha$ in terms of the total angular acceleration $\alpha = 3GM_\oplus/2r^3$ and inertia $I = mr^2$. Consequently, the angular momentum **L** of an on-board gyroscope will change $\tau = d_t\mathbf{L} = -\mathbf{L} \times \Omega_g$ so that the geodetic precession in the orbital plane accumulates at the rate (de Sitter 1916; Straumann 1984)

$$\mathbf{\Omega}_g = -\frac{3}{2}\mathbf{v} \times \nabla\phi = -\frac{3}{2}\mathbf{\omega}\mathbf{r} \cdot \nabla\phi = -\frac{3}{2}\frac{GM_\oplus}{c^2 r}\mathbf{\omega}. \quad (6)$$

The final functional form (Eq. 6) is in concert with general relativity. Accordingly the axis of an on-board gyroscope is calculated to turn in the orbital plane by -6606.3 mas per sidereal year in agreement with measurements -6601.8 ± 18.3 mas/year (Everitt et al. 2011). During one sidereal year GP-B completes 5383.4 orbits at $r$ = 7027.4 km corresponding to the period $\omega^{-1} = 97.7$ minutes (NASA GP-B Fact Sheet, 2005).

The orbiter will sense further excess in energy density due to the revolving Earth whose angular momentum $\mathbf{L}_\oplus = I_\oplus \boldsymbol{\omega}_\oplus$, where $I_\oplus$ is inertia and $\omega_\oplus$ is angular frequency. The excess density relative to the free space stems from the divergence-free part of the gravitational potential, i.e., vector potential $\mathbf{A}_\oplus = G\mathbf{L}_\oplus \times \mathbf{r}/c^2 r^3$ (Straumann 1984). It generates a rotational field $\mathbf{B}_\oplus = \nabla \times \mathbf{A}_\oplus$ whose cross product with $\mathbf{L}$ will cause a torque $\boldsymbol{\tau} = d_t\mathbf{L} = -\mathbf{L} \times (\nabla \times \mathbf{A}_\oplus) = -\mathbf{L} \times \boldsymbol{\Omega}_{fr}$ that will turn the on-board gyroscope away from an initial sighting to a distant star. This frame-dragging precession, also familiar from gravitomagnetism (Ciufolini & Wheeler 1995; Ruggiero & Tartaglia 2002; Veto 2010), accumulates in the equatorial plane of Earth at the rate (Lense & Thirring 1918; Schiff 1960; Straumann 1984)

$$\boldsymbol{\Omega}_{fr} = -\nabla \times \mathbf{A}_\oplus = -\frac{GI_\oplus}{c^2 r^3}\nabla \times (\mathbf{r} \times \boldsymbol{\omega}_\oplus) \\ = -\frac{GI_\oplus}{c^2 r^3}\left(\boldsymbol{\omega}_\oplus - \frac{3\mathbf{r}}{r^2}\mathbf{r} \cdot \boldsymbol{\omega}_\oplus\right) \quad (7)$$

obtained in the same ways as a dipole field (Feynman, Leighton & Sands 1964). The functional form is also in concert with general relativity. Accordingly the calculated average rate of precession for a GP-B gyroscope -37.4 mas away from the initial sighting to IM Pegasi during one sidereal year is in agreement with observations (Everitt et al. 2011). The actual readings of an on-board array of superconducting gyroscopes were I: $-41.3 \pm 24.6$, II: $-16.1 \pm 29.7$, III: $-25 \pm 12.1$, IV: $-49.3 \pm 11.4$ mas/year whose average $-37.2 \pm 7.2$ mas/year. Since GP-B orbits over the poles, its gyroscopes will turn in the equatorial plane at the average rate $\langle\Omega_{fr}\rangle = -GI_\oplus \cos(\delta)/2c^2 r^3$ away from the initial direction pointing to a star at a declination $\delta$.

## 4 DISCUSSION

The principle of least action in its original form can be written to relate a local density to the universal energy density. The local excess manifests itself, e.g., as perihelion, geodetic and Lense-Thirring precessions. The calculated numerical values agree with the GB-P probed precessions and are in concert with general relativity. This outcome is of course anticipated because both the least-action principle and general relativity give geodesics as solutions to their respective equations of motions. Already early on the refractive index was used to relate diverse densities to the density of free space (Mahoney 1994). Indeed, the least-time equation of motion given in terms of refractive index bears resemblance to the mathematical models of spacetime referred to as the metrics (Schwarzschild 1916; Reissner 1916; Nordström 1918; Kerr 1963). As usual, the refractive index may also express emission of bound quanta as well as absorption of free quanta (Feynman 1964) to denote the system's evolution as a result of net efflux or influx of quanta to its surroundings.

Although stationary paths by the least-time principle in its original form and general relativity parallel each other, differences will accumulate along evolutionary trajectories. The principle of least action can cope with changes in density, i.e., evolution (Kaila & Annila 2008; Annila & Salthe 2010), whereas general relativity and other metric theories that comply with a group of symmetry (e.g. the Poincaré group), are according to Noether's theorem constrained to invariance, i.e., to account only for stationary states (Noether 1918; Birkhoff 1924; Weinberg 1995). For example, when light from a distant supernova propagates through the expanding, hence diluting Universe, intensity of the explosion will fall inversely proportional to the square of the increasing luminosity distance and proportional to the frequency that shifts to red due to the dilution. These two factors will not yield a straight line but a curve, when magnitude versus logarithm of redshift is plotted (Annila 2011a). Likewise, light will not merely bend by gravitational attraction but curve more because the photon will shift its frequency to sweep equal arcs in equal intervals of time when passing from the density of free space through a local density (Annila 2011a). Moreover, the density will change substantially when an orbit extends from enclosing a local mass $M_o$ at a radius $r = \frac{1}{2}a_o t^2$, where $a_o = GM_o/r^2$, to a cosmic perimeter where the acceleration $a = GM/R^2 = 2c^2/R$ results from the total mass $M$ of the Universe of radius $R$ (Annila 2009). Therefore, the density experienced by orbiters, such as gas molecules well beyond the luminous edge of a galaxy, will govern their velocity $v = r/t$ according to the balance $mv^2/r = GM_o/r^2$ that transcribes by insertion of $r = \frac{1}{2}at^2$ to Tully-Fisher relation $v^4 = aGM_o/2 = cHGM_o$, where Hubble parameter $H = 1/T$ (Hubble 1929; Weinberg 1972). These results demonstrate that the principle of least time presents a general account of natural processes whereas the principle of equivalence provides a particular relation between gravitation and acceleration.

The principle of least action á la Maupertuis was recognized early on, but it was soon shunned, presumably because the powerful imperative delineates not only computable stationary paths but also path-dependent processes from one state to another (Annila 2011b). The non-holonomic character of nature does not appeal to one who prefers certainty. It turns out that only deterministic processes, i.e., those without alternative trajectories, or stationary motions on closed orbits can be calculated. In other words, the future can be "predicted" when there are no alternatives or when the process is reversible, i.e., without the notion of time's arrow.

The principle of least action regards everything in tangible forms of quanta. Hence also a flow of time is embodied by a flow of quanta from the system to its surroundings or vice versa (Tuisku, Pernu & Annila 2009; Annila 2010). Accordingly for a spatial coordinate to exist, it must embody non-vanishing density of quanta. This thermodynamic tenet means, e.g., that a clock ticks faster in the free space than in a higher potential because the local density cannot as readily accept the dissipated quanta. Accordingly, the clock ticks slower when in prograde motion with a revolving density because the experienced field generated by the vector potential is higher and hence the surroundings cannot as readily accept dissipation as when the clock is in retrograde motion (Mashhoon 1999). The thermodynamic tenet assigns dissipation to non-inertial motions and thereby implies that no body with mass can accelerate up to the speed of light.

It is not a new thought that photons embody the vacuum. The similar functional forms of Coulomb potential and gravitational scalar potential $\phi$ prompted already Oliver Heaviside to consider a gravitational vector potential **A** as the generator of a rotational field (Heaviside 1893). Also conservation in a form of free space gauge $\partial_t\phi + c^2\nabla\cdot\mathbf{A} = 0$ implies gravity with scalar and vector character. When sources are present, the balance is given by Eq. 1. In the context of electromagnetism the least-time balance between changes in the kinetic energy and scalar and vector potentials is usually known as Poynting's theorem (Tuisku, Pernu & Annila 2009). As usual, differentials of the scalar and vector potentials give rise to electric and magnetic fields. However, we emphasize that the gravitational scalar and vector potentials were not invoked here by analogy with electromagnetism but resulted from the principle of least action.

The maximum entropy partition of photons in balance with matter manifests itself in cosmic background radiation that complies with Planck's law. Also electromagnetic characteristics of the free space $\varepsilon_o$ and $\mu_o$ in the mass-energy equivalence $E = mc^2 = m/\varepsilon_o\mu_o$ suggest that photons embody not only electromagnetic field but also the density of space. Moreover, the vacuum's non-zero energy density manifests itself in Casimir effect (Casimir & Polder 1948). Indeed the vacuum does eject photons in the dynamic Casimir effect (Wilson et al., 2011). Also the double-slit experiment is easily comprehensible when a projectile is understood to induce perturbations to the vacuum density which will subsequently go through the slits as well, and interfere with the particle in propagation to produce an interference pattern. The Aharanov-Bohm variant of the double-slit experiment, in turn, demonstrates how an applied vector potential will increase the vacuum density and thereby affect the propagation of induced perturbations (Aharonov & Bohm 1959).

Today, the idea of luminiferous ether as a medium for light has been discarded, but it is still worth to consider that the photons themselves embody the vacuum. At first sight this proposed form of a physical vacuum may appear absurd because we do not observe light when an object falls down whereas an accelerating charge will unmistakably emit photons. On the other hand, to restore the fallen object up in its initial state, we will have to consume free energy that ultimately originates from insolation. Could it be that the net neutral body, when changing from one state to another, i.e., accelerating is emitting (or absorbing) not one but two photons of opposite polarization? Whence so, no light can be seen, but the photon pair will still carry energy density to the surrounding free space from the contracting space that is confined between the object and its attracting target.

Surely, the notion of vacuum embodied by photons is also contained in modern physics, however, only when the photons are deemed as virtual. For example, the clearly perceptible electric and magnetic fields are currently considered as being composed of virtual photons. Curiously, in certain experiments the virtual photons are pictured to "transform" to the real photons (Wilson et al., 2011) although such an account would seem to question the conservation of quanta. Here we ask ourselves, why to resort to virtual rather than real photons when describing how electromagnetic and gravitational potentials come about. A charged particle will induce the photons of the surrounding vacuum to depart from their random phase distribution as well as from their uniform distribution of density. Accordingly, a net neutral body will induce in its vicinity a density gradient of photons but without mutual coherence hence without electromagnetic field character. Thus the photon-embodied vacuum will raise a real-time response to any local perturbation by re-adjusting its density and phase. The photon-embodied vacuum communicates gravitational effects due to the entire Universe. These inertial forces are real but often deemed as fictitious forces (Veto 2011).

Admittedly, the photon-embodied vacuum has been disdained due to difficulties in understanding how mass and charge of a particle relate to each other. For example, proton and neutron have nearly equal masses but their charges and magnetic moments differ largely. This particular puzzle, however, can be solved when particles are described as actions (Annila 2010). Then the mass of a particle can be understood to relate to the projection of the corresponding curved geodesic on the straight contours of freely propagating surrounding actions. The sign and magnitude of a charge and magnetic moment, in turn, accumulate from the geodesic's sense and degree of chirality.

The proposed photon-embodied physical vacuum that gives rise to both gravitational and electromagnetic fields, sheds light also on recent experiments where inertial effects come into sight when a superconducting ring is accelerated (Tajmar 2007). The superconducting characteristics imply a highly stationary state that entrains also the density of surrounding photons which are here understood to embody both the electromagnetic and gravitational potentials (Lano 1996). The ring in a normal state does not trap the surrounding vacuum of photons hence its inertia is not sensed to the same degree by probes in the vicinity. Moreover, the innate relation between electromagnetic and gravitational fields via the photon-embodied physical vacuum may clarify the experimental tribulations encountered with the array of superconducting GP-B gyroscopes considering their conceivable interactions.

When a system changes its state for another, for example by accelerating, at least one quantum must either be absorbed or emitted. Hence also the surrounding energy density must restructure to supply the absorbed quanta or to accommodate the emitted quanta to satisfy conservation. The resistance in restructuring, ultimately involving the entire Universe, is known as inertia. The all-around propagating energy density couples everything with everything else in accord with Mach's principle (Bondi & Samuel 1996; Einstein 1923). The local motions are affected via the physical vacuum that tends by means of photon propagation to be in balance with the entire Universe. To ascribe inertia to the photon-embodied physical vacuum may seem incompatible with general relativity. Yet, a refusal by such reasoning may not be compelling because observations, as exemplified here, can also be rationalized by the least-time principle. The portrayal of vacuum as physical is not against Einstein's thoughts either. On the contrary, he reasoned that inertia originates in a kind of interaction between bodies (Einstein 1923) and wrote (Einstein 1920): *To deny the ether is ultimately to assume that empty space has no physical qualities whatever. The fundamental facts of mechanics do not harmonize with this view.*

**ACKNOWLEDGEMENTS**
I thank Mikael Koskela for insightful comments and corrections.